\documentclass[aps,showpacs,twocolumn,superscriptaddress]{revtex4}
\usepackage{amsmath}
\usepackage{amssymb}
\usepackage{epsfig}
\usepackage{bm}
\usepackage[english,russian]{babel}
\usepackage{indentfirst}

\usepackage{graphicx,amsmath}
\makeatletter
\renewcommand{\@biblabel}[1]{#1.}
\makeatother
\graphicspath{{Figures/}}

\begin{document}
\title{QUASIPARTICLES OF STRONGLY CORRELATED FERMI LIQUIDS
AT HIGH TEMPERATURES AND IN HIGH MAGNETIC FIELDS}
\author{\bf V. R. Shaginyan}\email{vrshag@thd.pnpi.spb.ru}
\affiliation{Petersburg Nuclear Physics Institute, RAS, Gatchina,
188300, Russia}

\begin{abstract}
Strongly correlated Fermi systems are among the most intriguing,
best experimentally studied and fundamental systems in physics.
There is, however, lack of theoretical understanding in this field
of physics. The ideas based on the concepts like Kondo lattice and
involving quantum and thermal fluctuations at a quantum critical
point have been used to explain the unusual physics. Alas, being
suggested to describe one property, these approaches fail to
explain the others. This means a real crisis in theory suggesting
that there is a hidden fundamental law of nature. It turns out that
the hidden fundamental law is well forgotten old one directly
related to the Landau---Migdal quasiparticles, while the basic
properties and the scaling behavior of the strongly correlated
systems can be described within the framework of the fermion
condensation quantum phase transition (FCQPT). The phase transition
comprises the extended quasiparticle paradigm that allows us to
explain the non-Fermi liquid (NFL) behavior observed in these
systems. In contrast to the Landau paradigm stating that the
quasiparticle effective mass is a constant, the effective mass of
new quasiparticles strongly depends on temperature, magnetic field,
pressure, and other parameters. Our observations are in good
agreement with experimental facts and show that FCQPT is
responsible for the observed NFL behavior and quasiparticles
survive both high temperatures and high magnetic fields.
\end{abstract}

 \maketitle

\section{Introduction} \label{INTR}

A. B. Migdal's contribution to modern theoretical physics is very
impressive. His endowment in and deep understanding of different
domains of physics, including the nuclear and many-body physics,
based on Fermi-liquid approach, is outstanding. In Migdal seminal
papers a solid base for studying strongly interacting Fermi systems
and phase transitions occurring in them has been established
\cite{mig1,mig2}. Migdal's daring ideas of phase transitions
related to $\pi$ condensation in nuclei and neutron stars
\cite{mig2} inspired a theory of fermion condensation that has
permitted to construct a new class of Fermi liquids, new
quasiparticles and new type of merging of single-particle levels of
both finite and infinite Fermi systems like nuclear, atomic and
solid state systems \cite{ks,ksk,vol,volovik2,merg}. The new class
of Fermi liquids is represented by strongly correlated Fermi
systems where enormous number of experimental facts are collected.
Understanding the physics of these systems stimulates intensive
studies of the possible manifestation of fermion condensation in
other areas, as it has happened in the case of metal
superconductivity, whose ideas were successfully used in describing
atomic nuclei \cite{mig1} and in a possible explanation of the
origin of the mass of elementary particles. Therefore, we expect
that the ideas associated with the new fermion condensation quantum
phase transition \cite{prep} in one area of research stimulates
intensive studies of the possible manifestation of such a
transition in other areas.

Strongly correlated Fermi systems represented by heavy fermion (HF)
metals and quasi-two-dimensional $^3$He are among the most
intriguing, best experimentally studied and fundamental systems in
physics, which until very recently have lacked theoretical
explanations \cite{prep}. These are also a field never far from
applications in synthesis of novel materials for cryogenics, rare
earth magnets and applied superconductivity. The properties of
these materials differ dramatically from those of ordinary Fermi
systems \cite{prep,ste,varma,vojta,voj,obz,col11}. Their behavior
is so unusual that the traditional Landau quasiparticles paradigm
does not apply to it. The paradigm states that the properties is
determined by quasiparticles whose dispersion is characterized by
the effective mass $M^*$ which is independent of temperature $T$,
the number density $x$, magnetic field $B$ and other external
parameters. The above systems are, however, in defiance of
theoretical understanding. The ideas based on the concepts (like
Kondo lattice involving quantum and thermal fluctuations at a
quantum critical point (QCP) have been used to explain the unusual
physics of these systems known as non-Fermi liquid (NFL) behavior.
Alas, being suggested to describe one property, these approaches
fail to explain the others. This means a real crisis in theory
suggesting that there is a hidden fundamental law of nature, which
remains to be recognized. It is widely believed that utterly new
concepts are required to describe the underlying physics. There is
a fundamental question: how many concepts do we need to describe
the above physical mechanisms? This cannot be answered on purely
experimental or theoretical grounds. Rather, we have to use both of
them. For instance, in the case of metals with heavy fermions, the
strong correlation of electrons leads to a renormalization of the
effective mass of quasiparticles, which may exceed the ordinary,
"bare", mass by several orders of magnitude or even become
infinitely large at temperatures $T\to0$. Moreover, the effective
mass strongly depends on the temperature, pressure, or applied
magnetic field. Such metals exhibit NFL behavior and unusual power
laws of the temperature dependence of the thermodynamic properties
at low temperatures.

The Landau theory of the Fermi liquid has remarkable results in
describing a multitude of properties of the electron liquid in
ordinary metals, Fermi liquids of the $^3$He type and nuclear
liquid \cite{landau,lanl1,migdal,PinNoz}. The theory is based on
the assumption that elementary excitations determine the physics at
low temperatures. These excitations behave as quasiparticles, have
a certain effective mass, and, judging by their basic properties,
belong to the class of quasiparticles of a weakly interacting Fermi
gas. Hence, the effective mass $M^*$ is independent of the
temperature, pressure, and magnetic field strength and is a
parameter of the theory. The Landau Fermi liquid (LFL) theory fails
to explain the results of experimental observations related to the
dependence of $M^*$ on the temperature $T$, magnetic field $B$,
pressure, etc.; this has led to the conclusion that quasiparticles
do not survive in strongly correlated Fermi systems and that the
heavy electron does not retain its identity as a quasiparticle
excitation, see e.g. \cite{col11,col2,col1}.

The unusual properties and NFL behavior observed in high-$T_c$
superconductors, HF metals and 2D Fermi systems are assumed to be
determined by various magnetic quantum phase transitions
\cite{ste,varma,vojta,voj,obz,col11,col2,geg1}. Indeed, when
reasoning by analogy with respect to the second order phase
transitions, one can assume that a phase transition responsible for
the NFL behavior taking place up to lowest accessible temperatures
is located at $T=0$. Since a quantum phase transition occurs at
$T=0$, the control parameters are the composition, electron (hole)
number density $x$, pressure, magnetic field strength $B$, etc. A
quantum phase transition occurs at a quantum critical point, which
separates the ordered phase that emerges as a result of quantum
phase transition from the disordered phase. It is usually assumed
that magnetic (e.g., ferromagnetic and antiferromagnetic) quantum
phase transitions are responsible for the NFL behavior. The
critical point of such a phase transition can be shifted to
absolute zero by varying the above parameters.

Universal behavior can be expected only if the system under
consideration is very close to a quantum critical point, e.g., when
the correlation length is much longer than the microscopic length
scale, and critical quantum and thermal fluctuations determine the
anomalous contribution to the thermodynamic functions of strongly
correlated Fermi system. Quantum phase transitions of this type are
so widespread \cite{varma,vojta,voj,col11,col2} that we call them
ordinary quantum phase transitions \cite{shag3}. In this case, the
physics of the phenomenon is determined by thermal and quantum
fluctuations of the critical state, while quasiparticle excitations
are destroyed by these fluctuations. Conventional arguments that
quasiparticles in strongly correlated Fermi liquids "get heavy and
die" at a quantum critical point commonly employ the well-known
formula based on the assumptions that the $z$-factor (the
quasiparticle weight in the single-particle state) vanishes at the
points of second-order phase transitions \cite{col1}. However, it
has been shown that this scenario is problematic \cite{khodb,x18}.

The fluctuations in the order parameter developing an infinite
correlation length and the absence of quasiparticle excitations are
considered as the main reason for the NFL behavior of heavy-fermion
metals, 2D fermion systems and high-$T_c$ superconductors
\cite{vojta,voj,col11,col2,col3}. This approach faces certain
difficulties, however. Critical behavior in experiments with metals
containing heavy fermions is observed at high temperatures
comparable to the effective Fermi temperature $T_k$. For instance,
the thermal expansion coefficient $\alpha(T)$, which is a linear
function of temperature for normal LFL, $\alpha(T)\propto T$,
demonstrates the $\sqrt{T}$ temperature dependence in measurements
involving CeNi$_2$Ge$_2$ as the temperature varies by two orders of
magnitude (as it decreases from 6 K to at least 50 mK) \cite{geg1}.
Such behavior can hardly be explained within the framework of the
critical point fluctuation theory. Obviously, such a situation is
possible only as $T\to0$, when the critical fluctuations make the
leading contribution to the entropy and when the correlation length
is much longer than the microscopic length scale. At a certain
temperature $T_k$, this macroscopically large correlation length
must be destroyed by ordinary thermal fluctuations and the
corresponding universal behavior must disappear.

In the rest of this paper, we show that the fermion condensation
quantum phase transition (FCQPT) \cite{prep} is indeed responsible
for the observed fascinating NFL behavior of strongly correlated
Fermi systems and quasiparticles survive both high temperatures and
high magnetic fields. In Section \ref{ehq}, we give a detailed
consideration of experimental evidences in favor of existence of
quasiparticles, and formulate both a scaling behavior of strongly
correlated Fermi systems and the extended quasiparticle paradigm.
Then in Section \ref{FLFC1}, we consider the properties of Landau
Fermi liquid. In Section \ref{POM_M} we demonstrate that the Landau
equation for the effective is not a phenomenological one and can be
derived using the methods of Density Functional Theory. Thus, we
establish the extended quasiparticle paradigm. FCQPT and a phase
diagram of heavy fermion system located in the vicinity of FCQPT
are investigated in Section \ref{FLFC}. We propose that the phase
diagram of systems located near FCQPT are strongly influenced by
control parameters such as a chemical pressure, pressure or
magnetic field. We find that under the application of the chemical
pressure (positive/negative) QCP is destroyed or converted into a
quantum critical line, correspondingly. In Section \ref{HMF} we
establish that heavy fermion quasiparticles do exist in a very wide
range of both temperatures $T$ and magnetic fields $B$. Finally, in
Section \ref{sum} our results are summarized and discussed.

\section{Experimental hints at scaling behavior and quasiparticles} \label{ehq}

There is a difficulty in explaining the restoration of the LFL
behavior under the application of magnetic field $B$, as observed
in HF metals and in high-$T_c$ superconductors \cite{ste,geg,cyr}.
For the LFL state as $T\to0$, the electric resistivity
$\rho(T)=\rho_0+AT^2$, the heat capacity $C(T)=\gamma_0T$, and the
magnetic susceptibility $\chi=const$. It turns out that the
coefficient $A(B)$, the Sommerfeld coefficient
$\gamma_0(B)=C/T\propto M^*$, and the magnetic susceptibility
$\chi(B)$ depend on the magnetic field strength B such that
$A(B)\propto\gamma_0^2(B)$ and $A(B)\propto\chi^2(B)$, which
implies that the Kadowaki---Woods relation $K=A(B)/\gamma_0^2(B)$
\cite{kadw} is $B$-independent and is preserved \cite{geg}. Such
universal behavior, quite natural when quasiparticles with the
effective mass $M^*$ playing the main role, can hardly be explained
within the framework of approach that presupposes the absence of
quasiparticles, which is characteristic of ordinary quantum phase
transitions in the vicinity of QCP. Indeed, there is no reason to
expect that $\gamma_0$, $\chi$ and $A$ are affected by the
fluctuations in a correlated fashion.

For instance, the Kadowaki---Woods relation does not agree with the
spin density wave scenario \cite{geg} and with the results of
research in quantum criticality based on the renormalization-group
approach \cite{mill}. Moreover, measurements of charge and heat
transfer have shown that the Wiedemann---Franz law holds in some
high-$T_c$ superconductors \cite{cyr,cyr1} and HF metals
\cite{pag,pag2,ronn1,ronn}. All this suggests that quasiparticles
do exist in such metals, and this conclusion is also corroborated
by photoemission spectroscopy results \cite{koral,fujim}.

The inability to explain the behavior of heavy-fermion metals while
staying within the framework of theories based on ordinary quantum
phase transitions implies that another important concept introduced
by Landau, the order parameter, also ceases to operate. Thus, we
are left without the most fundamental principles of many-body
quantum physics \cite{landau,lanl1,PinNoz}, and many interesting
phenomena associated with the NFL behavior of strongly correlated
Fermi systems remain unexplained.

\begin{figure} [! ht]
\begin{center}
\vspace*{-0.2cm}
\includegraphics [width=0.47\textwidth]{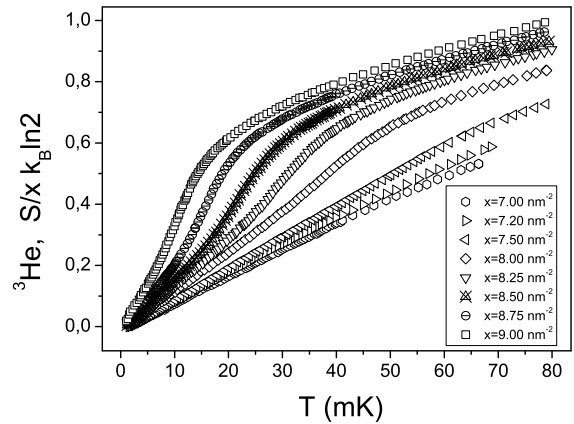}
\end{center}
\vspace*{-0.3cm} \caption{The entropy of 2D $\rm^3He$ as a function
of the density $x$ versus $T$ \cite{3he}. The density is depicted
in the legend.}\label{SHE}
\end{figure}

\begin{figure} [! ht]
\begin{center}
\vspace*{-0.2cm}
\includegraphics [width=0.47\textwidth]{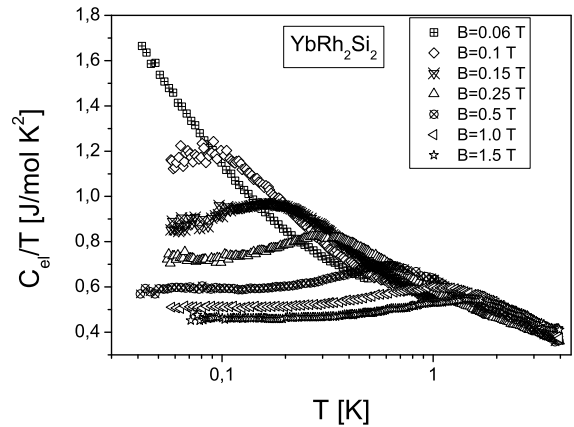}
\end{center}
\vspace*{-0.3cm} \caption{Electronic specific heat of $\rm
YbRh_2Si_2$, $C/T$, versus temperature $T$ as a function of
magnetic field $B$ \cite{oesb} shown in the legend.}\label{YBRHSI}
\end{figure}

\begin{figure} [! ht]
\begin{center}
\vspace*{-0.2cm}
\includegraphics [width=0.47\textwidth]{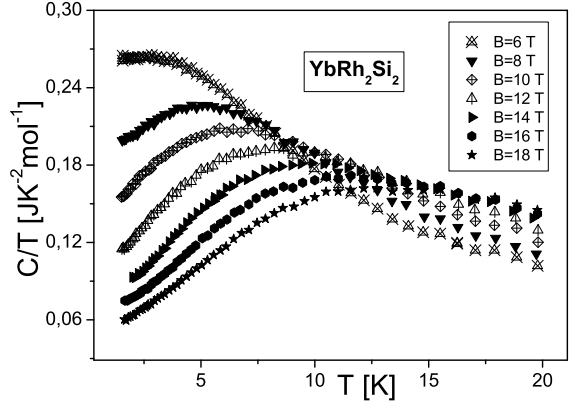}
\end{center}
\vspace*{-0.3cm} \caption{The normalized effective mass
$M^*_N=M^*/M^*_M$ versus normalized temperature $T_N=T/T_M$.
$M^*_N$ is extracted from the measurements of the specific heat
$C/T$ on $\rm YbRh_2Si_2$ \cite{steg1} shown in Fig. \ref{MHF}. The
values of the field $B$ is listed in the legend.}\label{MHF}
\end{figure}

The NFL behavior manifests itself in the power-law behavior of the
physical quantities of strongly correlated Fermi systems located
close to their QCPs, with exponents different from those of a Fermi
liquid
\cite{ste,varma,vojta,voj,obz,col11,col2,geg1,oesb,steg1,oesbs}. It
is common belief that the main output of theory is the explanation
of these exponents which are at least depended on the magnetic
character of QCP and dimensionality of the system. On the other
hand, the observed behavior of the thermodynamic properties cannot
be captured by these exponents as seen from Figs. \ref{SHE}
---\ref{MHF}. The behavior of the entropy $S(T)$ of two-dimensional
(2D) $\rm^3He$ \cite{3he} shown in Fig. \ref{SHE} is positively
different from that described by a simple function $a_1T^{a_2}$
where $a_1$ is a constant and $a_2$ is the exponent. It is seen
from Fig. \ref{SHE} that at the low densities $x\simeq 7$ $\rm
nm^{-2}$ the entropy demonstrates the LFL behavior characterized by
a linear function of $T$ with $a_2=1$. The behavior becomes quite
different at the higher densities at which $S(T)$ has an inflection
point. Obviously, at the inflection point $S(T)$ cannot be fit by
the simple function $a_1T^{a_2}$.

As seen from Fig. \ref{YBRHSI}, the specific heat $C/T$ measured on
$\rm YbRh_2Si_2$ \cite{oesb} exhibits a behavior that is to be
described as a function of both temperature $T$ and magnetic $B$
field rather than by a single exponent. One can see that at low
temperatures $C/T$ demonstrates the LFL behavior which is changed
by the transition regime at which $C/T$ reaches its maximum and
finally $C/T$ decays into NFL behavior as a function of $T$ at
fixed $B$. It is clearly seen from Fig. \ref{YBRHSI} that, in
particularly in the transition regime, these exponents may have
little physical significance.

Figure \ref{MHF} displays measurements of $C/T$ on $\rm YbRh_2Si_2$
\cite{steg1} in wide range of both temperature $T$ and magnetic
field $B$ variations. It is seen that both the NFL behavior and the
range extends at least up to twenty Kelvins and up to 18 T. Figure
\ref{MHF} demonstrates that the LFL state at low temperatures and
NFL one at higher temperatures are separated by the transition
regime at which $C/T$ reaches its maximum value $M^*_M(B)$. Thus,
we again conclude that the observed NFL behavior exhibiting the
maximum and extending to high temperatures can be hardly explained
in the framework of the ordinary quantum phase transitions
interpreting the exponents.

\begin{figure} [! ht]
\begin{center}
\vspace*{-0.2cm}
\includegraphics [width=0.47\textwidth]{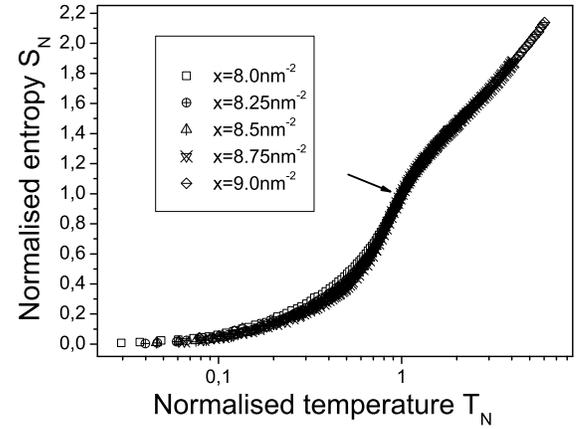}
\end{center}
\vspace*{-0.3cm} \caption{The normalized entropy $S_N$, extracted
from the experimental facts displayed in Fig. \ref{SHE}, as a
function of $x$ shown in the legend versus normalized $T_N$. The
arrow shows the inflection point.}\label{SHEN}
\end{figure}

In order to show that the behavior of $S$ and $C/T$  displayed in
Figs. \ref{SHE}---\ref{MHF} is of generic character, we remember
that in the vicinity of QCP it is helpful to use "internal" scales
to measure the effective mass $M^*\propto C/T$ and temperature $T$
\cite{prep,dft373,dftjtp}. The internal scales of the thermodynamic
functions such as $S$ or $C/T$ are related to "peculiar points"
like the inflection or maximum. Since the entropy has no maxima,
its normalization is to be performed in the inflection point taking
place at $T=T_{inf}$. Note that $T_{inf}$ is a function of $x$, it
is seen from Fig. \ref{SHE} that the inflection point moves towards
lower temperatures at elevating $x$. The normalized entropy $S_N$
as a function of the normalized temperature $T_N=T/T_{inf}=y$ is
reported in Fig. \ref{SHEN}.
We normalize the entropy by its value at the inflection point
$S_N(y)=S(y)/S(1)$. As seen from Fig. \ref{SHEN}, the normalization
revels the scaling behavior of $S_N$, that is the curves at
different temperatures and densities $x$ merge into a single one in
terms of the variable $y$. We have excluded the experimental data
taken at  $x\leq 8 \,{\rm nm^{-2}}$ since the corresponding curves
do not contain the inflection points. It is seen from Fig.
\ref{SHEN} that $S_N(y)$ extracted from the measurements is not a
linear function of $y$, as would be for a LFL, and shows the
scaling behavior over three decades in normalized temperature $y$.

\begin{figure} [! ht]
\begin{center}
\vspace*{-0.2cm}
\includegraphics [width=0.47\textwidth]{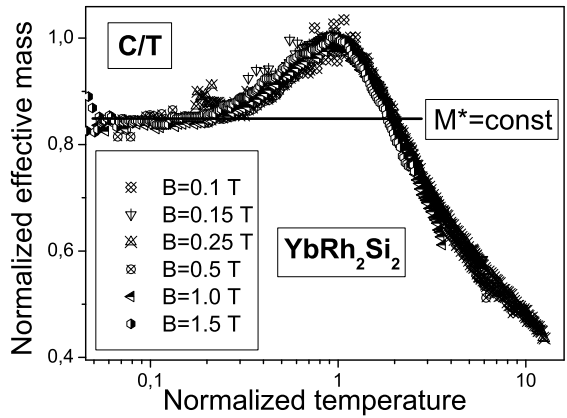}
\end{center}
\vspace*{-0.3cm} \caption{The normalized effective mass $M^*_N$
versus normalized temperature $T_N$.  $M^*_N$ is extracted from the
measurements of the specific heat $C/T$ on $\rm YbRh_2Si_2$ in
magnetic fields $B$ \cite{oesb} listed in the legend. Constant
effective mass $M^*_L$ inherent in normal Landau Fermi liquids is
depicted by the solid line.}\label{YBRHSIN}
\end{figure}

As seen from Fig. \ref{YBRHSI}, a maximum structure in $C/T\propto
M^*_M$ at temperature $T_M$ appears under the application of
magnetic field $B$ and $T_M$ shifts to higher $T$ as $B$ is
increased. The value of the Sommerfeld coefficient $C/T=\gamma_0$
is saturated towards lower temperatures decreasing at elevated
magnetic field. To obtain the normalized effective mass $M^*_N$, we
use $M^*_M$ and $T_M$ as "internal" scales: The maximum structure
in $C/T$ was used to normalize $C/T$, and $T$ was normalized by
$T_M$. In Fig. \ref{YBRHSIN} the normalized effective mass
$M^*_N=M^*/M^*_M$ as a function of normalized temperature
$T_N=T/T_M$ is shown by geometrical figures. Note that we have
excluded the experimental data taken in magnetic field $B=0.06$ T.
In that case, $T_M\to0$ and the corresponding $T_M$ and $M^*_M$ are
unavailable.
It is seen that the LFL state and NFL one are separated by the
transition regime at which $M^*_N$ reaches its maximum value.
Figure \ref{YBRHSIN} reveals the scaling behavior of the normalized
experimental curves: The curves at different magnetic fields $B$
merge into a single one in terms of the normalized variable
$y=T/T_M$. As seen from Fig. \ref{YBRHSIN}, the normalized
effective mass $M^*_N(y)$ extracted from the measurements is not a
constant, as would be for a LFL, and shows the scaling behavior
over three decades in normalized temperature $y$. It is seen from
Figs. \ref{YBRHSI} and \ref{YBRHSIN} that the NFL behavior and the
associated scaling extend at least to temperatures up to few
Kelvins.

\begin{figure} [! ht]
\begin{center}
\vspace*{-0.2cm}
\includegraphics [width=0.47\textwidth]{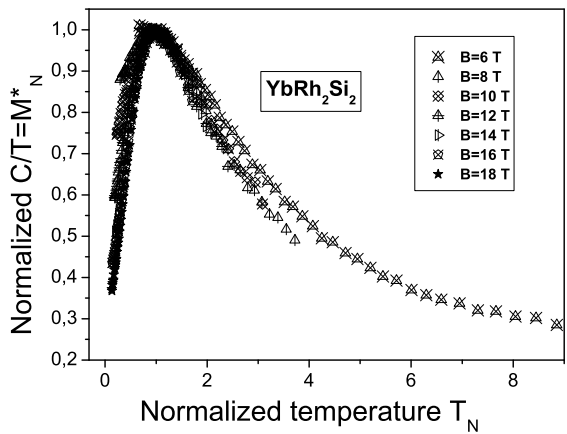}
\end{center}
\vspace*{-0.3cm} \caption{The normalized effective mass
$M^*_N=M^*/M^*_M$ versus normalized temperature $T_N=T/T_M$.
$M^*_N$ is extracted from the measurements of the specific heat
$C/T$ on $\rm YbRh_2Si_2$ \cite{steg1} shown in Fig. \ref{MHF}. The
values of the field $B$ is listed in the legend.}\label{MNHF}
\end{figure}

In order to get a deep insight in the behavior of $C/T$ displayed
in Fig. \ref{MHF}, we again use the internal scales as it was done
when constructing $M^*_N$ displayed in Fig. \ref{YBRHSIN}. As a
result, Fig. \ref{MNHF} reveals the scaling behavior of the
normalized experimental curves: The curves at different magnetic
fields $B$ merge into a single one in terms of the normalized
variable $y=T/T_M$. As seen from Fig. \ref{MNHF}, the normalized
effective mass $M^*_N(y)$ is not a constant, as would be for a LFL,
and shows the scaling behavior over two decades in normalized
temperature $y$. It is seen from Figs. \ref{MHF} and \ref{MNHF}
that the NFL behavior and the associated scaling extend at least
both to high temperatures up to twenty Kelvins and magnetic fields
up to 18 T.

Taking into account the scaling behavior revealed in Figs.
\ref{SHEN}---\ref{MNHF}, we conclude that a challenging problem for
theories considering the NFL behavior of strongly correlated Fermi
liquids is to explain both the scaling and the shape of the
thermodynamic functions like $S_N$ and $M^*_N$. While the theories
calculating only the exponents that characterize $S_N$ and $M^*_N$
at $y\gg 1$ deal with a part of the observed facts related to the
problem and overlook, for example, consideration of the transition
and LFL regimes. Another part of the problem is the remarkably
large ranges of temperature, magnetic field and the density over
which the NFL behavior and the scaling are observed. Concepts based
on the Kondo lattice and scenarios where fluctuations in the order
parameter is sufficiently big and the correlation time is
sufficiently large to develop the NFL behavior can hardly match up
such high temperatures and explain the observed scaling behavior.

As we will see below, the large temperature ranges are precursors
of new quasiparticles, and  it is the scaling behavior of the
normalized effective mass that allows us to explain the
thermodynamic of HF metals at the transition and NFL regimes.
Taking into account the simple behavior shown in Figs.
\ref{SHEN}---\ref{MNHF}, we ask the question: what theoretical
concepts can replace the Fermi-liquid paradigm with the notion of
the effective mass in cases where Fermi-liquid theory breaks down?
To date such a concept is not available \cite{vojta}. Therefore, we
focus on the concept of FCQPT preserving quasiparticles and
intimately related to the unlimited growth of $M^*$. It was shown
that it is capable of revealing the scaling behavior of the
effective mass and delivering an adequate theoretical explanation
of a vast majority of experimental results in different HF metals
\cite{prep}. In contrast to the Landau paradigm based on the
assumption that $M^*$ is a constant as shown by the solid line in
Fig. \ref{YBRHSIN}, in FCQPT approach the effective mass $M^*$ of
new quasiparticles strongly depends on $T$, $x$, $B$ etc.
Therefore, in accord with numerous experimental facts the extended
quasiparticles paradigm is to be introduced. The main point here is
that the well-defined quasiparticles determine as before the
thermodynamic, relaxation and transport properties of strongly
correlated Fermi-systems in large temperature ranges, while $M^*$
becomes a function of $T$, $x$, $B$ etc. The FCQPT approach had
been already successfully applied to describe the thermodynamic
properties of such different strongly correlated systems as $^3$He
on the one hand and complicated HF compounds on the other
\cite{prep,obz,khodb,prl3he}.

Since we are concentrated on properties that are non-sensitive to
the detailed structure of the system we avoid difficulties
associated with the anisotropy generated by the crystal lattice of
solids, its special features, defects, etc., We study the universal
behavior of strongly correlated Fermi-systems located near their
QCP at low temperatures using the model of a homogeneous HF liquid
\cite{dft373,dftjtp}. The model is quite meaningful because we
consider the scaling behavior exhibited by these materials at low
temperatures \cite{prep}. The scaling properties of the normalized
effective mass that characterizes them, are determined by momentum
transfers that are small compared to momenta of the order of the
reciprocal lattice length. The high momentum contributions can
therefore be ignored by substituting the lattice for the jelly
model. While the values of the scales like the maximum $M^*_M$ of
the effective mass and $T_M$ at which $M^*_M$ takes place are
determined by a wide range of momenta and thus these scales are
controlled by the specific properties of the system. On the other
hand, the dependencies of these scales on magnetic fields,
temperature, the density of system, pressure etc are again
controlled by small momenta and can be analyzed within the model of
HF liquid.

\section{Normal Fermi liquids}\label{FLFC1}

One of the most complex problems of modern condensed matter physics
is the problem of the structure and properties of Fermi systems with
large inter particle coupling constants. Theory of Fermi liquids,
later called "normal", was first proposed by Landau as a means for
solving such problems by introducing the concept of quasiparticles
and amplitudes that characterize the effective quasiparticle
interaction \cite{landau, lanl1}. The Landau theory can be regarded
as an effective low-energy theory with the high-energy degrees of
freedom eliminated by introducing amplitudes that determine the
quasiparticle interaction instead of the strong inter particle
interaction. The stability of the ground state of the Landau Fermi
liquid is determined by the Pomeranchuk stability conditions:
stability is violated when at least one Landau amplitude becomes
negative and reaches its critical value \cite{landau,lanl1,pom}. We
note that the new phase in which stability is restored can also be
described, in principle, by the LFL theory.

We begin by recalling the main ideas of the LFL theory
\cite{landau, lanl1,PinNoz}. The theory is based on the
quasiparticle paradigm, which states that quasiparticles are
elementary weakly excited states of Fermi liquids and are therefore
specific excitations that determine the low-temperature
thermodynamic and transport properties of Fermi liquids. In the
case of the electron liquid, the quasiparticles are characterized
by the electron quantum numbers and the effective mass $M ^*$. The
ground state energy of the system is a functional of the
quasiparticle occupation numbers (or the quasiparticle distribution
function) $n({\bf p},T)$, and the same is true of the free energy
$F(n({\bf p},T))$, the entropy $S(n({\bf p},T))$, and other
thermodynamic functions. We can find the distribution function from
the minimum condition for the free energy $F=E-TS$ (here and in
what follows $k_B=\hbar=1$)
\begin{equation}\label{FL1} \frac{\delta(F-\mu N)}{\delta n({\bf p}, T)}=\varepsilon({\bf
p}, T) -\mu (T)-T\ln\frac{1-n({\bf p},T)}{n({\bf p},T)}=0.
\end{equation}
Here $\mu$ is the chemical potential fixing the number density
\begin{equation}\label{NUMX}
x=\int n({\bf p},T)\frac{d{\bf p}}{(2\pi)^3},
\end{equation} and
\begin{equation} \varepsilon({\bf p},T) \ = \ \frac {\delta
E(n({\bf p},T))} {\delta n({\bf p},T)}\,\label{FL2} \end{equation}
is the quasiparticle energy. This energy is a functional of $n({\bf
p},T)$, in the same way as the energy $E$ is: $\varepsilon({\bf
p},T,n)$. The entropy $S(n({\bf p},T))$ related to quasiparticles
is given by the well-known expression \cite{landau,lanl1}
\begin{eqnarray}
S(n({\bf p},T))&=& -2\int[n({\bf p},T) \ln (n({\bf p},T))+(1-n({\bf p},T))\nonumber \\
&\times&\ln (1-n({\bf p},T))]\frac{d{\bf p}}{(2\pi) ^3},\label{FL3}
\end{eqnarray}
which follows from combinatorial reasoning. Equation \eqref{FL1} is
usually written in the standard form of the Fermi---Dirac
distribution,
\begin{equation} n({\bf p},T)=
\left\{1+\exp\left[\frac{(\varepsilon({\bf p},T)-\mu)}
{T}\right]\right\}^{-1}.\label{FL4} \end{equation} At $T\to 0 $,
(\ref{FL1}) and (\ref{FL4}) have the standard solution
$n(p,T\to0)\to\theta(p_f-p)$ if the derivative $\partial
\varepsilon(p\simeq p_{\rm F})/\partial p$ is finite and positive.
Here $p_{\rm F}$ is the Fermi momentum and $\theta (p_{\rm F}-p)$
is the step function. The single particle energy can be
approximated as $\varepsilon(p\simeq p_{\rm F})-\mu\simeq p_{\rm
F}(p-p_{\rm F})/M^*_L$, and $M ^*_L$ inversely proportional to the
derivative is the effective mass of the Landau quasiparticle,
\begin{equation}
\frac1{M^*_L}=\frac1p\, \frac{d\varepsilon(p,T=0)}{dp}|_{p=p_{\rm
F}}\ \label{FL5}.\end{equation} In turn, the effective mass $M^*_L$
is related to the bare electron mass $m$ by the well-known Landau
equation \cite{landau,lanl1,PinNoz} \begin{eqnarray}\label{LANDM}
\frac{1}{M^*_L} &=& \frac{1}{m}+\sum_{\sigma_1}\int \frac{{\bf
p}_F{\bf p_1}}{p_{\rm F}^3} F_{\sigma,\sigma_1}({\bf p_{\rm
F}},{\bf p}_1) \nonumber \\ &\times & \frac{\partial
n_{\sigma_1}({\bf p}_1,T)}{\partial {p}_1} \frac{d{\bf
p}_1}{(2\pi)^3}.
\end{eqnarray}
where $F_{\sigma,\sigma_1}({\bf p_{\rm F}},{\bf p}_1)$ is the
Landau amplitude, which depends on the momenta ${\bf p_{\rm F}}$
and ${\bf p}$ and the spins $\sigma$. For simplicity, we ignore the
spin dependence of the effective mass, because $M^*_L$ is almost
completely spin-independent in the case of a homogeneous liquid and
weak magnetic fields. The Landau amplitude $F$ is given by
\begin{equation}\label{AMPL} F_{\sigma,\sigma_1}({\bf p},{\bf p}_1,n)
=\frac{\delta^2E(n)}{\delta n_{\sigma}({\bf p})\delta
n_{\sigma_1}({\bf p}_1)}.\end{equation} The stability of the ground
state of LFL is determined by the Pomeranchuk stability conditions:
stability is violated when at least one Landau amplitude becomes
negative and reaches its critical value \cite{lanl1,PinNoz,pom}
\begin{equation}\label{POM}
F^{a,s}_L=-(2L+1).
\end{equation}
Here $F^{a}_L$ and $F^{s}_L$ are the dimensionless spin-symmetric
and spin-antisymmetric Landau amplitudes, $L$ is the angular
momentum related to the corresponding Legendre polynomials $P_L$,
\begin{equation}\label{LEG}
F({\bf p\sigma},{\bf
p}_1\sigma_1)=\frac{1}{N}\sum^{\infty}_{L=0}P_L(\Theta)
\left[F^{a}_L\sigma,\sigma_1 +F^{s}_L\right].
\end{equation}
Here $\Theta$ is the angle between momenta ${\bf p}$ and ${\bf
p}_1$ and the density of states $N=M^*_Lp_{\rm F}/(2\pi^2)$. It
follows from Eq. \eqref{LANDM} that
\begin{equation}\label{EFFM1}
\frac{M^*_L}{m}=1+\frac{F^s_1}{3}.
\end{equation}
In accordance with the Pomeranchuk stability conditions it is seen
from Eq. \eqref{EFFM1} that $F^s_1>-3$, otherwise the effective
mass becomes negative leading to unstable state when it is
energetically favorable to excite quasiparticles near the Fermi
surface.

\section{Effective mass and scaling behavior}\label{POM_M}

It is common belief that the equations of the Section \ref{FLFC1}
are phenomenological and inapplicable to describe Fermi systems
characterized by the effective mass $M^*$ strongly dependent on
temperature, external magnetic fields $B$, pressure $P$ etc. To
derive the equation determining the effective mass, we consider the
model of a homogeneous HF liquid and employ the density functional
theory for superconductors (SCDFT) \cite{gross} which allows us to
consider $E$ as a functional of the occupations numbers $n({\bf
p})$ \cite{dft373,dft,dft269,sh}. As a result, the ground state
energy of the normal state $E$ becomes the functional of the
occupation numbers and the function of the number density $x$,
$E=E(n({\bf p}),x)$, while Eq. \eqref{FL2} gives the
single-particle spectrum. Upon differentiating both sides of Eq.
\eqref{FL2} with respect to ${\bf p}$ and after some algebra and
integration by parts, we obtain \cite{khodb,dft373,dft,dft269}
\begin{equation}\label{EM}
\frac{\partial\varepsilon({\bf p})}{\partial {\bf p}}=\frac{{\bf
p}}{m}+\int F({\bf p},{\bf p}_1,n)\frac{\partial n({\bf
p}_1)}{\partial{\bf p}_1}\frac{d{\bf p}_1}{(2\pi)^3}.
\end{equation}
To calculate the derivative $\partial\varepsilon({\bf p})/\partial
{\bf p}$, we employ the functional representation
\begin{eqnarray}\label{ENP}
E(n)&=&\int\frac{p^2}{2m}n({\bf p})\frac{d{\bf
p}}{(2\pi)^3}\nonumber
\\&+&\frac{1}{2}\int F({\bf p},{\bf p}_1,n)_{|_{n=0}}\,n({\bf p})n({\bf
p}_1)\frac{d{\bf p}d{\bf p}_1}{(2\pi)^6} +...
\end{eqnarray}
It is seen directly from Eq. \eqref{EM} that the effective mass is
given by the well-known Landau equation \begin{equation}\label{FLL}
\frac{1}{M^*} = \frac{1}{m}+\int \frac{{\bf p}_F{\bf p_1}}{p_{\rm
F}^3} F({\bf p_{\rm F}},{\bf p}_1,n)\frac{\partial n(p_1)}{\partial
p_1} \frac{d{\bf p}_1}{(2\pi)^3}.
\end{equation}
For simplicity, we ignore the spin dependencies. To calculate $M^*$
as a function of $T$, we construct the free energy $F=E-TS$, where
the entropy $S$ is given by Eq. \eqref{FL3}. Minimizing $F$ with
respect to $n({\bf p})$, we arrive at the Fermi---Dirac
distribution, Eq. \eqref{FL4}.  Due to the above derivation, we
conclude that Eqs. \eqref{EM} and \eqref{FLL} are exact ones and
allow us to calculate the behavior of both
$\partial\varepsilon({\bf p})/\partial {\bf p}$ and $M^*$ which now
is a function of temperature $T$, external magnetic field $B$,
number density $x$ and pressure $P$ rather than a constant. As we
will see it is this feature of $M^*$ that forms both the scaling
and the NFL behavior observed in measurements on HF metals.

In LFL theory it is assumed that $M^*_L$ is positive, finite  and
constant since the integral on the right hand side of Eq.
\eqref{FLL} represents a small correction in the case of normal
metals. As a result, the temperature-dependent corrections to $M
^*_L$, the quasiparticle energy $\varepsilon({\bf p})$ and other
quantities begin with the term proportional to $T^2$ in 3D systems
and with the term proportional to $T$ in 2D one \cite{chub}. The
effective mass is given by Eq. \eqref{LANDM}, and the specific heat
$C$ is \cite{landau}
\begin{equation}\label{HEAT}
C=\frac{2\pi^2NT}{3}=\gamma_0T=T\frac{\partial S}{\partial T},
\end{equation}
and the spin susceptibility
\begin{equation}\label{SPINS}
\chi=\frac{3\gamma_0\mu_B^2}{\pi^2(1+F^a_0)},
\end{equation}
where $\mu_B$ is the Bohr magneton and $\gamma_0\propto M^*_L$. In
the case of LFL the electrical resistivity at low $T$ is given by
\cite{PinNoz}
\begin{equation}\label{RESIS}
\rho(T)=\rho_0+AT^{2},
\end{equation}
where $\rho_0$ is the residual resistivity and $A$ is the
coefficient determining the charge transport. The coefficient $A$
is proportional to the quasiparticle-quasiparticle scattering
cross-section. Equation \eqref{RESIS} symbolizes and defines the
LFL behavior observed in normal metals.

Equation (\ref{FLL}) at $T=0$, combined with the fact that $n({\bf
p},T=0)$ becomes $\theta (p_{\rm F}-p) $, yields the well-known
result \cite{vollh,pfw,vollh1}
$$ \frac{M^*}{m}=\frac{1}{1-f_N^1/3}.$$
where $f_N^1=N_0f^1$, $N_0=mp_{\rm F}/(2\pi^2)$ is the density of
states of a free Fermi gas and $f^1(p_{\rm F},p_{\rm F})$ is the
$p$-wave component of the Landau interaction amplitude. Because
$x=p_{\rm F}^3/3\pi^2$ in the Landau Fermi-liquid theory, the
Landau interaction amplitude can be written as $f_N^1(p_{\rm
F},p_{\rm F})=f_N^1(x)$. Provided that at a certain critical point
$x_{\rm FC}$, the denominator $(1-f_N^1(x)/3)$ tends to zero, i.e.,
$(1-f_N^1(x)/3)\propto(x-x_{\rm FC})+a(x-x_{\rm FC})^2 + ...\to 0$,
we find that \cite{khod1,shag1}
\begin{equation}
\frac{M^*(x)}{m}\simeq a_1+\frac{a_2}{x-x_{\rm
FC}}\propto\frac{1}{r}. \label{FL7}\end{equation} where $a_1$ and
$a_2$ are constants and $r=(x-x_{\rm FC})/x_{\rm FC}$ is the
``distance'' from QCP $x_{\rm FC}$ at which $M^*(x\to x_{\rm
FC})\to\infty$. We note that the divergence of the effective mass
given by Eq. \eqref{FL7} does preserve the Pomeranchuk stability
conditions for $f_N^1$ is positive, see Eq. \eqref{POM}. Equations
\eqref{EFFM1} and \eqref{FL7} seem to be different but it is not
the case since $f_N^1\propto m$, while $F^s_1\propto M^*$ and Eq.
\eqref{EFFM1} represents an implicit formula for the effective
mass.

The behavior of $M^*(x)$ described by formula (\ref{FL7}) is in
good agreement with the results of experiments
\cite{cas1,skdk,skdk1} and calculations \cite{krot,sarm1,sarm2}. In
the case of electron systems, Eq. \eqref{FL7} holds for $x>x_{\rm
FC}$, while for 2D $^3$He we have $x<x_{\rm FC}$ so that always
$r>0$ \cite{ksk,ksz}. Such behavior of the effective mass is
observed in HF metals, which have a fairly flat and narrow
conductivity band corresponding to a large effective mass, with a
strong correlation and the effective Fermi temperature $T_k\sim
p_{\rm F}^2/M ^*(x)$ of the order of several dozen degrees kelvin
or even lower (e.g., see Ref. \cite{ste}).

\begin{figure} [! ht]
\begin{center}
\includegraphics [width=0.47\textwidth] {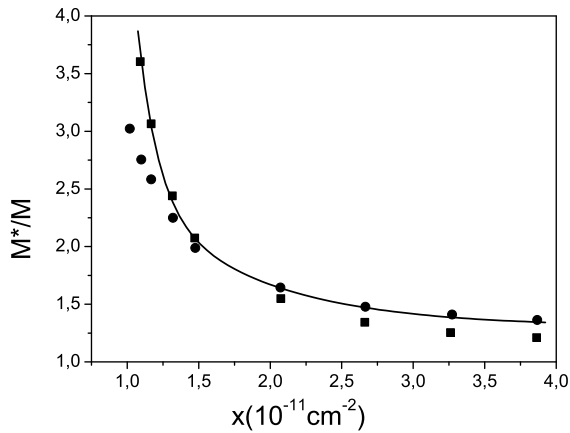}
\end{center}
\caption {The ratio $M ^*/M$ in a silicon MOSFET as a function of
the electron number density $x$. The squares mark the experimental
data on the Shubnikov-de Haas oscillations, and the data obtained
by applying a parallel magnetic field are marked by circles
\cite{skdk,skdk1,krav}. The solid line represents the function
(\ref{FL7}).} \label{Fig5}
\end{figure}

\begin{figure} [! ht]
\begin{center}
\includegraphics [width=0.47\textwidth] {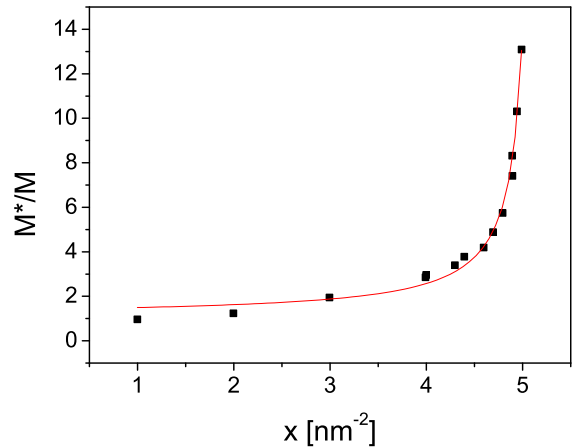}
\end{center}
\caption {The ratio $M ^*/M$ in 2D $^3$He as a function of the
density $x$ of the liquid, obtained from heat capacity and
magnetization measurements. The experimental data are marked by
squares \cite{cas1,cas}, and the solid line represents the function
given by Eq. \eqref{FL7}.} \label {Fig6}
\end{figure}

The effective mass as a function of the electron density $x$ in a
silicon MOSFET (Metal Oxide Semiconductor Field Effect Transistor),
approximated by Eq. (\ref{FL7}), is shown in Fig. \ref{Fig5}. The
parameters $a_1$, $a_2$ and $x_{\rm FC}$ are taken as fitting. We
see that Eq. \eqref{FL7} provides a good description of the
experimental results.
The divergence of the effective mass  $M^*(x)$ discovered in
measurements involving 2D $^3$He \cite{cas1,krav,cas} is
illustrated in Fig. \ref{Fig6}. Figures \ref{Fig5} and \ref{Fig6}
show that the description provided by Eq. \eqref{FL7} does not
depend on elementary Fermi particles constituting the system and is
in good agreement with the experimental data.
It is instructive to briefly explore the scaling behavior of $M^*$
in order to illustrate the ability of the quasiparticle extended
paradigm to capture the scaling behavior. Let us write the
quasiparticle distribution function as $n_1({\bf p})=n({\bf
p},T)-n({\bf p})$, with $n({\bf p})$ being the step function, and
Eq. \eqref{FLL} then becomes
\begin{equation}
\frac{1}{M^*(T)}=\frac{1}{M^*}+\int\frac{{\bf p}_F{\bf p_1}}{p_{\rm
F}^3}F({\bf p_{\rm F}},{\bf p}_1)\frac{\partial n_1(p_1,T)}
{\partial p_1}\frac{d{\bf p}_1}{(2\pi) ^3}. \label{LF1}
\end{equation}
At QCP $x\to x_{\rm FC}$, the effective mass $M^*(x)$ diverges and
Eq. \eqref{LF1} becomes homogeneous determining $M^*$ as a function
of temperature while the system exhibits the NFL behavior. If the
system is located before QCP, $M^*$ is finite, at low temperatures
the integral on the right hand side of Eq. \eqref{LF1} represents a
small correction to $1/M^*$ and  the system demonstrates the LFL
behavior seen in Figs. \ref{YBRHSI} and \ref{YBRHSIN}. The LFL
behavior assumes that the effective mass is independent of
temperature, $M^*(T)\simeq const$, as shown by the horizontal line
in Fig. \ref{YBRHSIN}. Obviously, the LFL behavior takes place only
if the second term on the right hand side of Eq. \eqref{LF1} is
small in comparison with the first one. Then, as temperature rises
the system enters the transition regime: $M^*$ grows, reaching its
maximum $M^*_M$ at $T=T_M$, with subsequent diminishing. As seen
from Fig. \ref{YBRHSIN}, near temperatures $T\geq T_M$ the last
"traces" of LFL regime disappear, the second term starts to
dominate, and again Eq. \eqref{LF1} becomes homogeneous, and the
NFL behavior is restored, manifesting itself in decreasing $M^*$ as
a function of $T$.

\section{Fermion condensation quantum phase transition}\label{FLFC}

As shown in Section \ref{POM_M}, the Pomeranchuk stability
conditions do not encompass all possible types of instabilities and
that at least one related to the divergence of the effective mass
given by Eq. \eqref{FL7} was overlooked \cite{ks}. This type of
instability corresponds to a situation where the effective mass, the
most important characteristic of quasiparticles, can become
infinitely large. As a result, the quasiparticle kinetic energy is
infinitely small near the Fermi surface and the quasiparticle
distribution function $n({\bf p})$ minimizing $E(n({\bf p}))$ is
determined by the potential energy. This leads to the formation of a
new class of strongly correlated Fermi liquids with FC
\cite{ks,ksk,vol_1,volovik2,vol}, separated from the normal Fermi
liquid by FCQPT \cite{shag3,ms,shb,sh1}.

It follows from (\ref{FL7}) that at  $T=0$ and as $r\to0$ the
effective mass diverges, $M^*(r)\to\infty$. Beyond the critical
point $x_{\rm FC}$, the distance $r$ becomes negative and,
correspondingly, so does the effective mass. To avoid an unstable
and physically meaningless state with a negative effective mass,
the system must undergo a quantum phase transition at the critical
point $x=x_{\rm FC}$, which, as we will see shortly, is FCQPT
\cite{ms,shb,sh1,shag3}. Because the kinetic energy of
quasiparticles that are near the Fermi surface is proportional to
the inverse effective mass, the potential energy of the
quasiparticles near the Fermi surface determines the ground-state
energy as $x\to x_{\rm FC}$. Hence, at $T=0$ a phase transition
reduces the energy of the system and transforms the quasiparticle
distribution function. Beyond QCP $x=x_{\rm FC}$, the quasiparticle
distribution is determined by the ordinary equation for a minimum
of the energy functional \cite{ks}:
\begin{equation} \frac{\delta E(n({\bf p}))}{\delta
n({\bf p})}=\varepsilon({\bf p})=\mu; \, p_i\leq p\leq p_{\rm F}.
\label{FL8}\end{equation}

\begin{figure} [! ht]
\begin{center}
\includegraphics [width=0.47\textwidth]{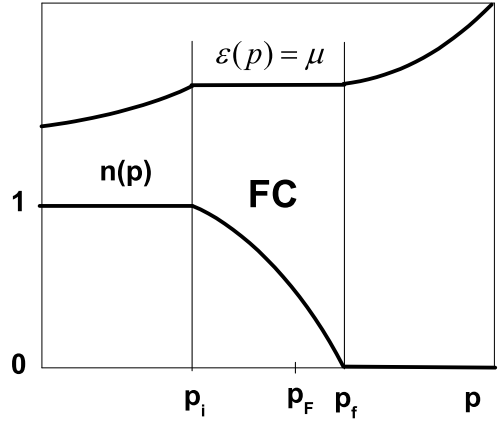}
\end{center}
\caption {The single-particle spectrum $\varepsilon(p)$ and the
quasiparticle distribution function $n_0(p)$. Because $n_0(p)$ is a
solution of Eq. (\ref{FL8}) at $T=0$, we have $n_0(p<p_i)=1$,
$0<n_0(p_i<p<p_{f})<1$, and $n_0(p>p_{f})=0$, while $\varepsilon
(p_i<p<p_{f})=\mu$ and the band becomes completely flat. The Fermi
momentum $p_{\rm F}$ satisfies the condition $p_i<p_{\rm
F}<p_{f}$.} \label{Fig1}
\end{figure}

Equation (\ref{FL8}) yields the quasiparticle distribution function
$n_0({\bf p}) $ that minimizes the ground-state energy $E$. This
function found from Eq. (\ref{FL8}) differs from the step function
in the interval from $p_i$ to $p_{\rm F}$, where $0<n_0({\bf
p})<1$, and coincides with the step function outside this interval.
In fact, Eq. \eqref{FL8} coincides with Eq. \eqref{FL2} provided
that the Fermi surface at $p=p_{\rm F}$ transforms into the Fermi
volume at $p_i\leq p\leq p_f$ suggesting that the single-particle
spectrum is absolutely ``flat'' within this interval. A possible
solution $n_0({\bf p})$ of Eq. (\ref{FL8}) and the corresponding
single-particle spectrum $\varepsilon({\bf p})$ are depicted in
Fig. \ref{Fig1}. Quasiparticles with momenta within the interval
$(p_f-p_i)$ have the same single-particle energies equal to the
chemical potential $\mu$ and form FC, while the distribution
$n_0({\bf p})$ describes the new state of the Fermi liquid with FC
\cite{ks,ksk,vol}. In contrast to the Landau, marginal, or
Luttinger Fermi liquids \cite{varma}, which exhibit the same
topological structure of the Green's function, in systems with FC,
where the Fermi surface spreads into a strip, the Green's function
belongs to a different topological class. The topological class of
the Fermi liquid is characterized by the invariant
\cite{volovik1,volovik2,vol}
\begin{equation} N=tr\oint_C\frac{dl}{2\pi i}G(i\omega,{\bf p})
\partial_lG^{-1}(i\omega,{\bf p})\label{FLVOL},\end{equation}
where ``tr'' denotes the trace over the spin indices of the Green's
function and the integral is taken along an arbitrary contour $C$
encircling the singularity of the Green's function. The invariant
$N$ in (\ref{FLVOL}) takes integer values even when the singularity
is not of the pole type, cannot vary continuously, and is conserved
in a transition from the Landau Fermi liquid to marginal liquids
and under small perturbations of the Green's function. As shown by
Volovik \cite{volovik1,volovik2,vol}, the situation is quite
different for systems with FC, where the invariant $N$ becomes a
half-integer and the system with FC transforms into an entirely new
class of Fermi liquids with its own topological structure.

\subsection*{Phase diagram of Fermi system with FCQPT}\label{FCPD}

We start with visualizing the main properties of FCQPT. To this
end, again consider SCDFT. SCDFT states that the thermodynamic
potential $\Phi$ is a universal functional of the number density
$n({\bf r})$ and the anomalous density (or the order parameter)
$\kappa({\bf r},{\bf r}_1)$, providing a variational principle to
determine the densities. At the superconducting transition
temperature $T_c$ a superconducting state undergoes the second
order phase transition. Our goal now is to construct a quantum
phase transition which evolves from the superconducting one, since,
as seen from Fig. \ref{Fig1}, the superconducting order parameter
$\kappa(p)=\sqrt{n(p)(1-n(p))}$ is finite over the region
$(p_{f}-p_i)$.

Let us assume that the coupling constant $\lambda_0$ of the
BCS-like pairing interaction \cite{bcs} vanishes, with
$\lambda_0\to0$ making vanish the superconducting gap at any finite
temperature. In that case, $T_c\to0$ and the superconducting state
takes place at $T=0$ while at finite temperatures there is a normal
state. This means that at $T=0$ the anomalous density
\begin{equation}\label{ANOM}\kappa({\bf
r},{\bf r_1})=\langle\Psi\uparrow({\bf r})\Psi\downarrow({\bf
r_1})\rangle\end{equation} is finite, while the superconducting gap
\begin{equation}\label{DEL}\Delta({\bf
r})=\lambda_0\int\kappa({\bf r},{\bf r_1})d{\bf r_1}\end{equation}
is infinitely small \cite{obz,shag1}. In Eq. \eqref{ANOM}, the
field operator $\Psi_{\sigma}({\bf r})$ annihilates an electron of
spin $\sigma, \sigma=\uparrow,\downarrow$ at the position ${\bf
r}$. For the sake of simplicity, we consider the model of
homogeneous HF liquid \cite{obz}. Then at $T=0$, the thermodynamic
potential $\Phi$ reduces to the ground state energy $E$ which turns
out to be a functional of the occupation number $n({\bf p})$ since
in that case the order parameter $\kappa({\bf p})=v({\bf p})u({\bf
p})=\sqrt{n({\bf p})(1-n({\bf p}))}$. Indeed,
\begin{equation} n({\bf p})=v^2({\bf
p}); \, \, \, \kappa({\bf p})=v({\bf p})u({\bf p}),\label{SC2}
\end{equation}
where $u({\bf p})$ and $v({\bf p})$ are normalized parameters such
that $v^2({\bf p})+u^2({\bf p})=1$ and $\kappa({\bf
p})=\sqrt{n({\bf p})(1-n({\bf p}))}$, see e.g. \cite{lanl1}.

Upon minimizing $E$ with respect to $n({\bf p})$, we obtain Eq.
\eqref{FL8}. As soon as Eq. \eqref{FL8} has nontrivial solution
$n_0({\bf p})$ then instead of the Fermi step, we have $0<n_0({\bf
p})<1$ in certain range of momenta $p_i\leq p\leq p_{\rm F}$ with
$\kappa({\bf p})=\sqrt{n_0({\bf p})(1-n_0({\bf p}))}$ being finite
in this range, while the single particle spectrum $\varepsilon({\bf
p})$ is flat. Thus, the step-like Fermi filling inevitably
undergoes restructuring and forms FC when Eq. \eqref{FL8} possesses
for the first time the nontrivial solution at $x=x_c$ which is QCP
of FCQPT. In that case, the range vanishes, $p_i\to p_{f}\to p_{\rm
F}$, and the effective mass $M^*$ diverges at QCP
\cite{obz,khodb,ks}
\begin{equation}\label{EFM}
\frac{1}{M^*(x\to x_c)}=\frac{1}{p_{\rm
F}}\frac{\partial\varepsilon({\bf p})}{\partial{\bf p}}|_{p\to
p_{\rm F};\,x\to x_c}\to 0.\end{equation} At any small but finite
temperature the anomalous density $\kappa$ (or the order parameter)
decays and this state undergoes the first order phase transition
and converts into a normal state characterized by the thermodynamic
potential $\Phi_0$. Indeed, at $T\to0$, the entropy $S=-\partial
\Phi_0/\partial T$ of the normal state is given by Eq. \eqref{FL3}.
It is seen from Eq. \eqref{FL3} that the normal state is
characterized by the temperature-independent entropy $S_0$
\cite{obz,khodb,yakov}. Since the entropy of the superconducting
ground state is zero, we conclude that the entropy is discontinuous
at the phase transition point, with its discontinuity $\delta
S=S_0$. Thus, the system undergoes the first order phase
transition. The heat $q$ of transition from the asymmetrical to the
symmetrical phase is $q=T_cS_0=0$ since $T_c=0$. Because of the
stability condition at the point of the first order phase
transition, we have $\Phi_0(n({\bf p}))=\Phi(\kappa({\bf p}))$.
Obviously the condition is satisfied since $q=0$.

At $T=0$, a quantum phase transition is driven by a nonthermal
control parameter, e.g., the number density $x$. As we have seen in
Section \ref{POM_M}, at QCP, $x=x_{\rm FC}$, the effective mass
diverges. It follows from Eq. \eqref{FL7} that beyond QCP, the
effective mass becomes negative. To avoid an unstable and
physically meaningless state with a negative effective mass, the
system undergoes FCQPT leading to the formation of FC.
\begin{figure} [! ht]
\begin{center}
\vspace*{-0.5cm}
\includegraphics [width=0.47\textwidth]{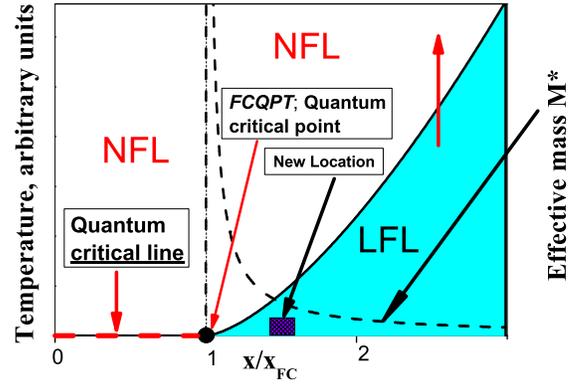}
\end{center}
\vspace*{-0.8cm} \caption{Schematic phase diagram of system with
FC. The number density $x$ is taken as the control parameter and
depicted as $x/x_{\rm FC}$. The dashed line shows $M^*(x/x_{\rm
FC})$ as the system approaches QCP, $x/x_{\rm FC}=1$, of FCQPT
which is denoted by the arrow. At $x/x_{\rm FC}>1$ and sufficiently
low temperatures, the system is in the LFL state as shown by the
shadow area. The new location of the system induced by the
application of the positive pressure is shown by both the solid
square and the arrow. The vertical arrow illustrates the system
moving in the LFL-NFL direction along $T$ at fixed control
parameter. At $T=0$ and beyond the critical point, $x/x_{\rm
FC}<1$, the system is at the quantum critical line depicted by the
dashed line and shown by the vertical arrow. The critical line is
characterized by the FC state with finite superconducting order
parameter $\kappa$. At any finite low temperature $T>T_c=0$,
$\kappa$ is destroyed, the system undergoes the first order phase
transition, possesses finite entropy $S_0$ and exhibits the NFL
behavior at any finite temperatures $T<T_f$.}\label{fig1}
\end{figure}

A schematic phase diagram of the system which is driven to the FC
state by variation of $x$ is reported in Fig. \ref{fig1}.  Upon
approaching the critical density $x_{\rm FC}$ the system remains in
the LFL region at sufficiently low temperatures as it is shown by
the shadow area. The temperature range of the shadow area shrinks
as the system approaches QCP, and $M^*(x/x_{\rm FC})$ diverges as
shown by the dashed line and Eq. \eqref{FL7}. At QCP $x_{\rm FC}$
shown by the arrow in Fig. \ref{fig1}, the system demonstrates the
NFL behavior down to the lowest temperatures. Beyond the critical
point at finite temperatures the behavior remains the NFL and is
determined by the temperature-independent entropy $S_0$
\cite{obz,yakov}. In that case at $T\to 0$, the system is again
demonstrating the NFL behavior and approaching a quantum critical
line (QCL) (shown by the vertical arrow and the dashed line in Fig.
\ref{fig1}) rather than a quantum critical point. Upon reaching the
quantum critical line from the above at $T\to0$ the system
undergoes the first order quantum phase transition, which is FCQPT
taking place at $T_c=0$. While at diminishing temperatures, the
systems located before QCP do not undergo a phase transition and
their behavior transits from NFL to LFL.

It is seen from Fig. \ref{fig1} that at finite temperatures there
is no boundary (or phase transition) between the states of systems
located before or behind QCP shown by the arrow. Therefore, at
elevated temperatures the properties of systems with $x/x_{\rm
FC}<1$ or with $x/x_{\rm FC}>1$ become indistinguishable. On the
other hand, at $T>0$ the NFL state above the critical line and in
the vicinity of QCP is strongly degenerate, therefore the
degeneracy stimulates the emergence of different phase transitions
lifting it and the NFL state can be captured by the other states
such as superconducting (for example, by the superconducting state
(SC) in $\rm CeCoIn_5$ \cite{shag1,yakov}) or by antiferromagnetic
(AF) state (e.g. AF one in $\rm YbRh_2Si_2$ \cite{dft373}) etc. The
diversity of phase transitions occurring at low temperatures is one
of the most spectacular features of the physics of many HF metals.
Within the scenario of ordinary quantum phase transitions, it is
hard to understand why these transitions are so different from one
another and their critical temperatures are so extremely small.
However, such diversity is endemic to systems with a FC
\cite{khodb}.

As mentioned above, the system located above QCL exhibits the NFL
behavior down to lowest temperatures unless it is captured by a
phase transition. The behavior exhibiting by the system located
above QCL is in accordance with the experimental observations that
study the evolution of QCP in $\rm YbRh_2Si_2$ under negative
chemical pressure induced with $\rm Ir/Ge$ substitution
\cite{stegnat,steg2}. As reported, the negative pressure leaves an
intermediate spin-liquid-type ground state over a finite magnetic
field range. We assume a simple model that the application of
negative pressure reduces $x$ and the system moves to a position
over QCL. That is the $\rm Ir/Ge$ substitution drives $\rm
YbRh_2Si_2$ to a paramagnetic state over the finite magnetic field
$B$ range, so that the electronic system of the HF metal is located
above QCL shown in Fig. \ref{fig1} and demonstrates the NFL
behavior. In that case magnetic field $B$ represents an additional
control parameter driving $\rm YbRh_2Si_2$ to the paramagnetic
state. We predict that at lowering temperatures, the electronic
system of $\rm YbRh_2Si_2$ is captured by a phase transition, since
the NFL state above QCL is strongly degenerate. At diminishing
temperatures, this degeneracy is to be removed by some phase
transition which likely can be detected by the LFL state
accompanying it.

Upon using nonthermal tuning parameters like the number density
$x$, the NFL behavior can be destroyed and the LFL one will be
restored. In our simple model, the application of positive pressure
$P$ makes $x$ grow removing the QCP of FCQPT shown in Fig.
\ref{fig1}, so that the electronic system of $\rm YbRh_2Si_2$ moves
into the shadow area characterized by the LFL behavior at low
temperatures. The new location of the system is shown by the arrow
pointing at the solid square. As a result, the application of
magnetic field $B\simeq B_{c0}$ does not drive the system to its
QCP with the divergent effective mass because the QCP is already
destroyed by the pressure. Here $B_{c0}$ is the critical magnetic
field that eliminates the corresponding AF order. At $B>B_{c0}$ and
raising temperatures, the system moving along the vertical arrow
transits from the LFL regime to the NFL one. As seen from Fig.
\ref{fig1}, in the NFL regime the system properties coincide with
that of the system at $P=0$. The observed behavior is in accord
with the experimental facts indicating the separation of the AF
order from QCP under the application of pressure in $\rm
YbRh_2Si_2$ \cite{stegjap}.

\section{High magnetic fields thermodynamics of heavy fermion
quasiparticles}\label{HMF}

As was mentioned in Section \ref{ehq}, one of the most interesting
and puzzling issues in the research of HF metals is the NFL behavior
taking place in wide range of both temperature $T$ and magnetic
field $B$ variations \cite{hmbt}. For example, measurements of the
specific heat on $\rm YbRh_2Si_2$ under the application of magnetic
field $B$ show that the range extends at least up to twenty Kelvins
as shown in Fig. \ref{MHF}. In Fig. \ref{MNHF} the normalized
effective mass $M^*_N$ as a function of normalized temperature $T_N$
is shown by geometrical figures. Figures  \ref{MHF} and \ref{MNHF}
reveal the remarkably large temperature ranges over which the NFL
behavior and scaling one are observed.

In this Section, we analyze the thermodynamic properties of $\rm
YbRh_2Si_2$ in both low and high magnetic fields. Our calculations
of the specific heat and magnetization allow us to conclude that
under the application of magnetic field the heavy-electron system
of $\rm YbRh_2Si_2$ evolves continuously without a metamagnetic
transition. At low temperatures and high magnetic fields $B\simeq
B^*$ the system is completely polarized and demonstrates the LFL
behavior, at elevated temperatures the HF behavior and related NFL
one are restored. The obtained results are in good agreement with
experimental facts as both temperature and magnetic field  vary by
more than two orders of magnitude.

We use the Landau equation \eqref{FLL} to study the behavior of the
effective mass $M^*(T,B)$ as a function of the temperature and the
magnetic field. For the model of homogeneous HF liquid at finite
temperatures and magnetic fields, Eq. \eqref{FLL} acquires the form
\cite{landau,prep}
\begin{eqnarray}
\nonumber \frac{1}{M^*_{\sigma}(T,
B)}&=&\frac{1}{m}+\sum_{\sigma_1}\int\frac{{\bf p}_F{\bf p}}{p_{\rm
F}^3}F_
{\sigma,\sigma_1}({\bf p_{\rm F}},{\bf p}) \\
&\times&\frac{\partial n_{\sigma_1} ({\bf
p},T,B)}{\partial{p}}\frac{d{\bf p}}{(2\pi)^3}. \label{HC1}
\end{eqnarray}
where $F_{\sigma,\sigma_1}({\bf p_{\rm F}},{\bf p})$ is the Landau
amplitude dependent on momenta $p_{\rm F}$, $p$ and spin $\sigma$.
For the sake of definiteness, we assume that the HF liquid is 3D
liquid. The quasiparticle distribution function can be expressed as
\begin{equation} n_{\sigma}({\bf p},T)=\left\{ 1+\exp
\left[\frac{(\varepsilon_{\sigma}({\bf
p},T)-\mu_{\sigma})}T\right]\right\} ^{-1},\label{HC2}
\end{equation}
where $\varepsilon_{\sigma}({\bf p},T)$ is the single-particle
spectrum. In our case, the chemical potential may depend on the
spin due to the Zeeman splitting $\mu_{\sigma}=\mu\pm \mu_BB$,
where $\mu_B$ is the Bohr magneton and $\mu$ is the chemical
potential preserving the number of particles.  We write the
quasiparticle distribution function as $n_{\sigma}({\bf p},T,B)
\equiv\delta n_{\sigma}({\bf p},T,B)+n_{\sigma}({\bf p},T=0,B=0)$.
Equation (\ref{HC1}) then becomes
\begin{eqnarray}
\nonumber &&\frac{1}{M^*(T,B)}=\frac{1}{M^*}+\frac{1}{p_{\rm
F}^2}\sum_{\sigma_1}
\int\frac{{\bf p}_F{\bf p_1}}{p_{\rm F}}\\
&\times&F_{\sigma,\sigma_1}({\bf p_{\rm F}},{\bf
p}_1)\frac{\partial\delta n_{\sigma_1}({\bf p}_1,T,B)}
{\partial{p}_1}\frac{d{\bf p}_1}{(2\pi) ^3}. \label{HC3}
\end{eqnarray}

Equation \eqref{HC3} is solved with a quite general form of Landau
interaction amplitude \cite{ckhz}. Choice of the amplitude is
dictated by the fact that the system has to be at the quantum
critical point of FCQPT, which means that the first two
$p$-derivatives of the single-particle spectrum $\varepsilon({\bf
p})$ should equal zero. Since the first derivative is proportional
to the reciprocal quasiparticle effective mass $1/M^*$, its zero
just signifies QCP of FCQPT. The second derivative must vanish;
otherwise $\varepsilon(p)-\mu$ has the same sign below and above
the Fermi surface, and the Landau state becomes unstable
\cite{khodb,prep}. Zeros of these two subsequent derivatives mean
that the spectrum $\varepsilon({\bf p})$ has an inflection point at
$p_{\rm F}$ so that the lowest term of its Taylor expansion is
proportional to $(p-p_{\rm F})^3$.

When the system is near FCQPT, it turns out that the normalized
solution of Eq. \eqref{HC3} $M^*_N(y)$ can be well approximated by
a simple universal interpolating function. The interpolation occurs
between the LFL ($M^*\propto a+ bT^2$) and NFL ($M^*\propto
T^{-2/3}$) regimes\cite{prep}
\begin{equation}M^*_N(y)\approx c_0\frac{1+c_1y^2}{1+c_2y^{8/3}}.
\label{UN2}
\end{equation}
Here $a$ and $b$ are constants, $c_0=(1+c_2)/(1+c_1)$, $c_1$ and
$c_2$ are fitting parameters, related to the Landau amplitude.
Magnetic field $B$ enters Eq. \eqref{HC2} as $B\mu_B/T$ making
$T_M\propto B\mu_B$ \cite{ckhz,prep}. We conclude that under the
application of magnetic field the variable
\begin{equation}\label{YTB}
y=T/T_M\propto \frac{T}{\mu_B(B-B_{c0})}
\end{equation}
remains the same and the normalized effective mass is again
governed by Eq. \eqref{UN2}. Here $B_{c0}$ is the critical magnetic
field driving both HF metal to its magnetic field tuned QCP and
corresponding N\'eel temperature to $T=0$. In some cases as in the
HF metal $\rm CeRu_2Si_2$, $B_{c0}=0$, see e.g. \cite{takah}, while
in $\rm YbRh_2Si_2$, $B_{c0}\simeq 0.06$ T \cite{geg} and in $\rm
CeCoIn_5$ the critical field is much larger $B_{c0}\simeq 5.0$ T
\cite{pag}. Therefore, in our simple model of the HF liquid
$B_{c0}$ is taken as a parameter. In what follows, we compute the
effective mass using Eq. \eqref{HC1} and employ Eq. \eqref{UN2} for
qualitative estimations of the considered values.
\begin{figure}[!ht]
\begin{center}
\includegraphics [width=0.48\textwidth]{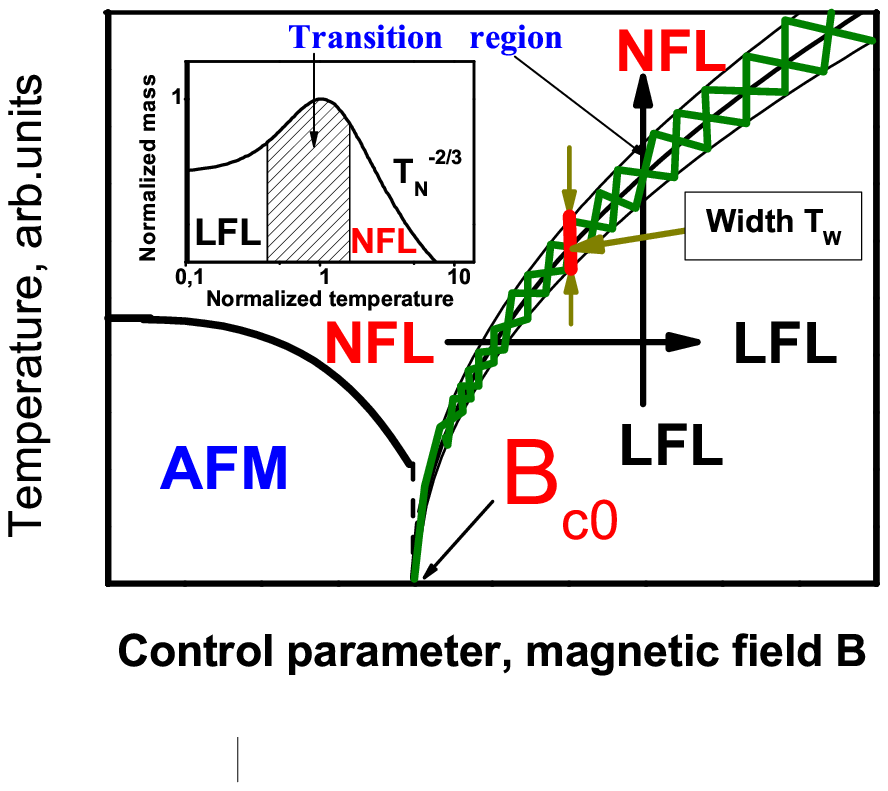}
\vspace*{-1.0cm}
\end{center}
\caption{Schematic phase diagram of $\rm YbRh_2Si_2$ based on Eq.
\eqref{UN2} with magnetic field as the control parameter $B$. The
vertical and horizontal arrows show LFL-NFL and NFL-LFL transitions
at fixed $B$ and fixed $T$ correspondingly. At $B <B_{c0}$ the
system is in antiferromagnetic (AFM) state. The width of the
transition regime $T_w\propto T$ is shown by the segment confined
by the two vertical arrows. Inset shows a schematic plot of the
normalized effective mass versus the normalized temperature.
Transition regime, where $M^*_N$ reaches its maximum value at
$y=T/T_M=1$, is shown by the hatched area both in the main panel
and in the inset.}\label{MT}
\end{figure}

We now are in position to construct the schematic phase diagram of
the HF metal $\rm YbRh_2Si_2$ at $B\ll B^*$. The pase diagram is
reported in Fig. \ref{MT}. The magnetic field $B$ plays a role of
the control parameter, driving the system towards its QCP. In our
case this QCP is of FCQPT type. The FCQPT peculiarity occurs at
$B=B_{c0}$, yielding new strongly degenerate state at $B<B_{c0}$.
To lift this degeneracy, the system forms either superconducting
(SC) or magnetically ordered (ferromagnetic (FM), antiferromagnetic
(AFM) etc) states \cite{prep}. In the case of $\rm YbRh_2Si_2$,
this state is AFM one \cite{steg1}. As it follows from Eqs.
\eqref{UN2} and \eqref{YTB} and seen from Fig. \ref{MT}, at $B\geq
B_{c0}$ the system is in either NFL or LFL states. At fixed
temperatures the increase of $B$ drives the system along the
horizontal arrow from NFL state to LFL one. On the contrary, at
fixed magnetic field and raising temperatures the system transits
along the vertical arrow from LFL state to NFL one. The inset to
Fig. \ref{MT} demonstrates the behavior of the normalized effective
mass $M^*_N=M^*/M^*_M$ versus normalized temperature $y=T/T_M$
following from Eq. \eqref{UN2}. The $T^{-2/3}$ regime is marked as
NFL one since (contrary to LFL case where the effective mass is
constant) the effective mass depends strongly on temperature. It is
seen that temperature region $y\sim 1$ signifies a transition
regime between the LFL behavior with almost constant effective mass
and NFL one, given by $T^{-2/3}$ dependence. Thus, temperatures
$T\simeq T_M$, shown by arrows in the inset and main panel, can be
regarded as a transition regime between LFL and NFL states. It is
seen from Eq. \eqref{YTB} that the width of the transition regime
$T_w\propto T$ is proportional to $(B-B_{c0})$. It is shown by the
segment between two vertical arrows in Fig \ref{MT}. These
theoretical results are in good agreement with the experimental
facts \cite{steg1,pnas}.

\begin{figure} [! ht]
\begin{center}
\includegraphics [width=0.47\textwidth]{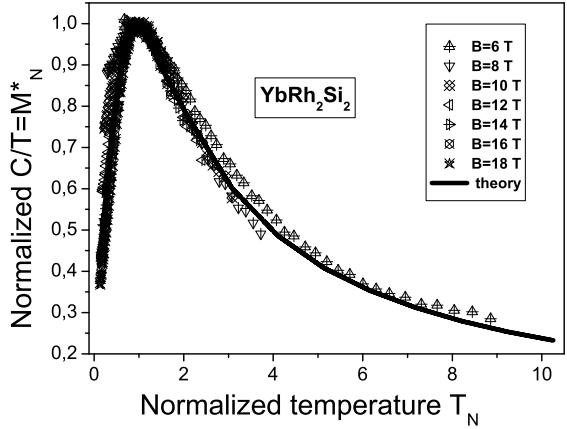}
\end{center}
\caption{The normalized effective mass $M^*_N=M^*/M^*_M$ versus
normalized temperature $T_N=T/T_M$.  $M^*_N$ is extracted from the
data shown in Fig. \ref{MHF}. Magnetic fields $B$ is listed in the
legend. Our calculations (made at $B\simeq B^*$ when the
quasiparticle band is fully polarized) are depicted by the solid
curve tracing the scaling behavior of $M^*_N$.}\label{MRM}
\end{figure}

In Fig. \ref{MRM}, our calculations of the normalized effective
mass $M^*_N(T_N)$ at fixed magnetic field $B^*$ making the system
fully polarized are shown by the solid line. Figure \ref{MRM}
reveals the scaling behavior of the normalized experimental curves
- the scaled curves at different magnetic fields $B$ merge into a
single one in terms of the normalized variable $y=T/T_M$. It is
seen from Fig. \ref{MRM} that our calculations are in accord with
facts \cite {steg1}: At elevated temperatures ($y\simeq 1$) the LFL
state first changes into the transition regime and then disrupts
into the NFL state.

\begin{figure} [! ht]
\begin{center}
\includegraphics [width=0.49\textwidth]{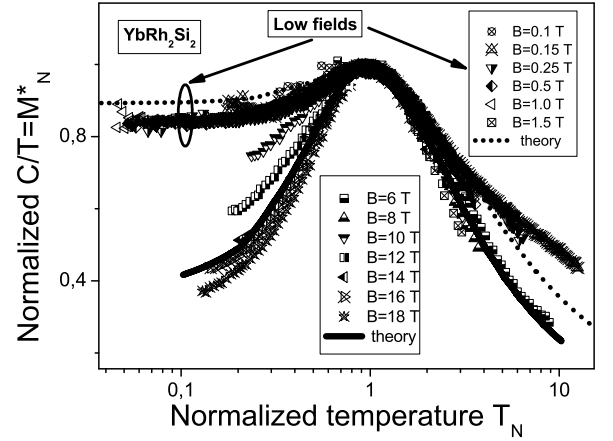}
\end{center}
\caption{Joint behavior of the normalized effective mass $M^*_N$ at
low (upper box symbols) and high (lower box symbols) magnetic
fields extracted from the specific heat ($C/T$) measurements of the
$\rm YbRh_2Si_2$ \cite{oesb}. Our low-field calculations are
depicted by the dotted line tracing the scaling behavior of
$M^*_N$. Our high-field calculations (solid line) are taken at
$B\sim B^*$ when the quasiparticle band becomes fully
polarized.}\label{f2}
\end{figure}

To get further insight into the behavior of the system at high
magnetic fields, in Fig. \ref{f2} we collect the curves
$M^*_N(T_N)$ both at low (symbols in the upper box in Fig.
\ref{f2}) and high (symbols in the lower box) magnetic fields $B$.
All curves have been extracted from the experimental facts
\cite{steg1,oesb}. It is seen that while at low fields the
low-temperature ends ($T_N \sim 0.1$) of the curves completely
merge, at high fields this is not the case. Moreover, the
low-temperature asymptotic value of $C/T=M^*_N$ at low fields is
around two times more then that at high fields. The physical reason
for low-field curves merging is that the effective mass does not
depend on spin variable so that the polarizations of subbands with
$\sigma_{\uparrow}$ and $\sigma_{\downarrow}$ are almost equal to
each other. This is reflected in our calculations, based on Eq.
\eqref{UN2} for low magnetic fields $B<<B^*$. The result is shown
by the dotted line in Fig. \ref{f2}.
It is also seen from Fig. \ref{f2} that all low-temperature
differences between high- and low field behavior of the normalized
effective mass disappear at high temperatures. In other words,
while at low temperatures the values of $M^*_N$ for low fields are
two times more then those for high fields, at temperatures
$T_N\geq1$ this difference disappear. It is seen that these high
temperatures lie about the transition region, marked by hatched
area in the inset to Fig. \ref{MT}. This means that two states (LFL
and NFL) separated by the transition region are clearly seen in
Fig. \ref{f2} displaying good agreement between our calculations
(dotted line for low fields and thick line at high fields) and the
experimental points (symbols).

Contrary to the low fields case, at high fields $B\sim B^*$
(symbols in the lower box of Fig. \ref{f2}), in the low temperature
LFL state the curves $M^*_N(T_N)$ do not merge into single one and
their values decrease as $B$ grows representing the full spin
polarization of the HF band at the highest reached magnetic fields.
As we have mentioned above, at temperature raising all effects of
spin polarization smear down, yielding the restoration NFL behavior
at $T\simeq \mu_BB$. Our high-field calculations (solid line in
Fig. \ref{f2}) reflect the latter fact and are also in good
agreement with experimental facts. In order not to overload Fig.
\ref{f2} with unnecessary details, we show the calculations only
for the case of the full spin polarization.
\begin{figure} [! ht]
\begin{center}
\includegraphics [width=0.49\textwidth]{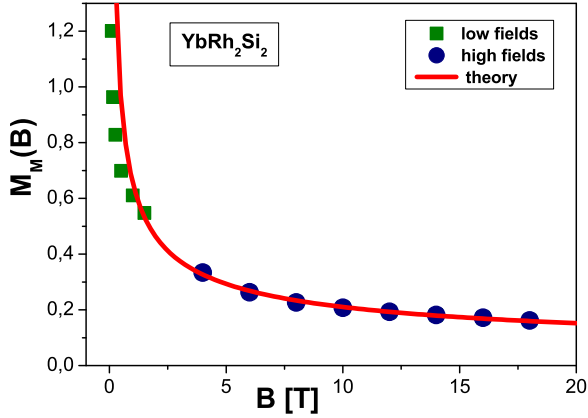}
\end{center}
\caption{The maxima $M^*_M(B)$ of the functions $C/T$ versus
magnetic field $B$ for $\rm YbRh_2Si_2$. The maxima taken at the
low fields \cite{oesb} are shown by the solid squares, the ones
taken at the high fields \cite{steg1} are depicted by the solid
circles. The solid curve is approximated by $M^*_M(B)\propto
d/\sqrt{B-B_{c0}}$, $d$ is a fitting parameter.}\label{MB}
\end{figure}
Figure \ref{MB} reports the maxima $M^*_M(B)$ of the functions
$C/T$ shown in Fig. \ref{MHF} versus $B$ depicted by the solid
circles and the solid squares, corresponding measurements on $\rm
YbRh_2Si_2$ taken at the low magnetic fields \cite{oesb}. The solid
line represents the approximation describing the maxima value of
$M^*(B)\propto \sqrt{B-B_{c0}}$ calculated within the framework of
FCQPT theory \cite{shag4,prep}. It is seen that our calculations
are in good agreement with the experimental facts over two decades
in magnetic field $B$. The agreement indicates that at $T\simeq
\mu_BB$ the transition regime takes place and the NFL behavior
restores at higher temperatures $T\sim 20$ K.

\begin{figure} [! ht]
\begin{center}
\includegraphics [width=0.47\textwidth]{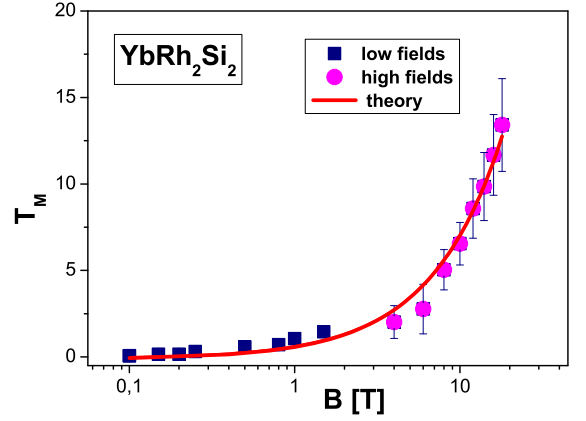}
\end {center}
\caption{The temperatures $T_M(B)$ at which the maxima of $C/T$ are
located versus magnetic field $B$ for $\rm YbRh_2Si_2$. The solid
squares denote $T_M$ taken in measurements at the low magnetic
fields \cite{oesb}. The solid circles denote $T_M$ at the high
magnetic fields \cite{steg1}. The solid line represents the
function $T_{M}\propto b(B-B_{c0})$, $b$ is a fitting parameter,
see Eq. \eqref{YTB}.} \label{TB}
\end{figure}

In Fig. \ref{TB}, the solid squares and circles denote temperatures
$T_N(B)$ at which the maxima of $C/T$ take place. To fit the
experimental data \cite{steg1,oesb} the function
$T_N(B)=b(B-B_{c0})$ defined by Eq. \eqref{YTB} is used. It is seen
from Fig. \ref{TB} that our calculations shown by the solid line
are in accord with experimental facts, and we conclude that the
transition regime of $\rm YbRh_2Si_2$ is restored at temperatures
$T\simeq \mu_B B$.
Consider now the magnetization $M(B,T)$ as a function of magnetic
field $B$ at fixed temperature $T=T_f$
\begin{equation}\label{CHIB}
M(B,T)=\int_0^B \chi(z,T)dz,
\end{equation}
where the magnetic susceptibility $\chi$ is given by
\cite{landau,lanl1}
\begin{equation}\label{CHI}
\chi(B,T)=\frac{\beta M^*(B,T)}{1+F_0^a}.
\end{equation}
Here, $\beta$ is a constant and $F_0^a$ is the Landau amplitude
related to the exchange interaction.

\begin{figure} [! ht]
\begin{center}
\includegraphics [width=0.49\textwidth]{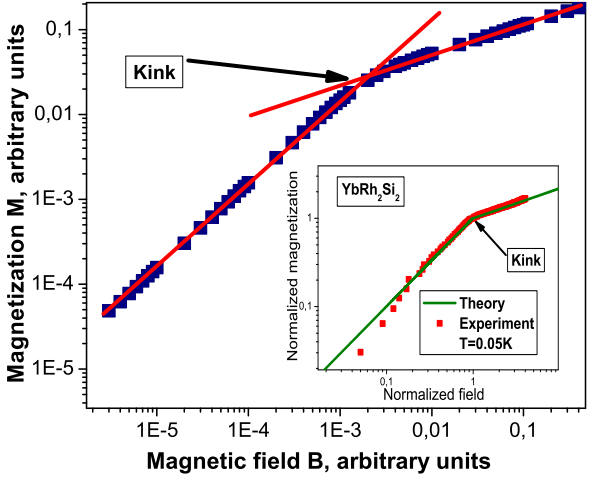}
\end{center}
\caption{The calculated magnetization $M(B)$ (symbols) and straight
lines which are guides for eye. The intersection of the straight
lines visualize the kink at the crossover region in the Fig.
\ref{MT}. The inset: The field dependencies of the normalized
magnetization $M$ of $\rm{ YbRu_2Si_2}$ \cite{oesbs} at $T=0.05$ K.
The kink (shown by the arrow) is clearly seen at the normalized
field $B_N=B/B_k\simeq 1$. The solid curve represents our
calculations.}\label{kink}
\end{figure}

\begin{figure} [! ht]
\begin{center}
\includegraphics [width=0.47\textwidth]{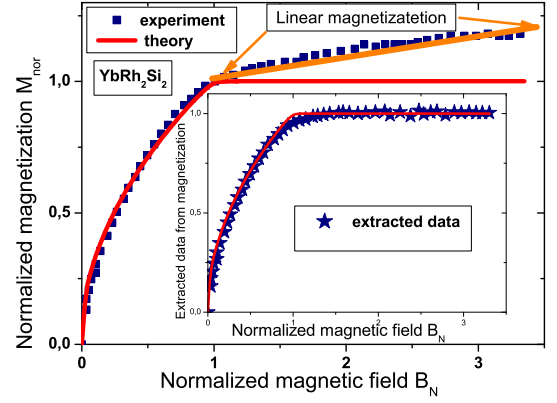}
\end {center}
\caption {The normalized magnetization $M_{nor}$ as a function of
the normalized magnetic field $B_N$. The calculated magnetization
is shown by the solid line, and the experimental facts \cite{steg1}
is represented by the solid squares. The linear dependence of
$M_{nor}$ on $B_N$ is shown by the arrows. The inset demonstrates
the experimental data with the subtracted linear dependence
depicted by the stars. Our calculations is shown by the solid
line.}\label{kink2}
\end{figure}

At low magnetic fields, $B\ll B^*$, our calculations show that the
magnetization exhibits a kink at some magnetic field $B=B_k$. The
experimental magnetization demonstrates the same behavior
\cite{steg,oesbs}. We use $B_k$ and $M(B_k)$ to normalize $B$ and
$M$ respectively. Due to the normalization the coefficients $\beta$
and $(1+F_0^a)$ drop out from the result, and $\chi\propto M^*$
\cite{prep}. So we can use Eq. \eqref{HC3} to calculate the
magnetic susceptibility $\chi$. The normalized magnetization
$M(B)/M(B_k)$ both extracted from experiment (symbols) and
calculated one (solid line), are reported in the inset to Fig.
\ref{kink}. It shows that our calculations are in good agreement
with the experimental facts, and all the data exhibit the kink
(shown by the arrow) at $B_N\simeq 1$ taking place as soon as the
system enters the transition region corresponding to the magnetic
fields where the horizontal arrow in Fig. \ref{MT} crosses the
hatched area. To illuminate the kink position, in the Fig.
\ref{kink} we present the $M(B)$ dependence in logarithmic -
logarithmic scale. In that case the straight lines show clearly the
change of the power dependence $M(B)$ at the kink.
At magnetic field $B\simeq B^*$ the quasiparticle band becomes
fully polarized and a new kink appears \cite{steg1,mal}, we call
this kink as the second one. Our calculations of the normalized
magnetization (line) and the experimental points (squares) are
shown in Fig. \ref{kink2}. In that case both the magnetization and
the field are normalized by the corresponding values at the second
kink.
In the Fig. \ref{kink2} we plot our calculated magnetization (in
the variables $B$ and $M$ normalized to those in the second kink
point) versus experimental one. The marvelous coincidence is seen
everywhere except the high-field part at $B_N\geq 1$ where the
normalized magnetization $M_{nor}$ exhibits a linear dependence on
$B_N$ depicted by the two arrows, while the calculated
magnetization is approximately constant at these $B_N$. Such a
behavior is the intrinsic shortcoming of the HF liquid model that
takes into account only heavy electrons and omit the other
conduction electrons that contribute to the magnetization
\cite{c1,c2}. Thus, we can consider the high-field (at $B_N>1$)
part of the magnetization as the contribution which does not
included in our theory. To separate this contribution from the
experimental magnetization curve, we (numerically) differentiate
it, then subtract constant part at $B_N>1$ and integrate back the
resulting curve. The coincidence between our calculations depicted
by the solid curve and processed experimental data shown by the
stars is reported in the inset to Fig. \ref{kink2}. As we can see
now, the coincidence between the theory and experiment is good in
the entire magnetic field domain. Taking into account the obtained
results displayed in Figs. \ref{f2}---\ref{kink2}, we conclude that
the HF system of $\rm{ YbRu_2Si_2}$ evolves continuously under the
application of magnetic fields. This observation is in agreement
with experimental observations \cite{prl}.

\section{Summary}\label{sum}

We have shown that the physics of systems with heavy fermions is
determined by the extended quasiparticle paradigm. In contrast to
the paradigm that the quasiparticle effective mass is a constant,
within the extended quasiparticle paradigm the effective mass of
new quasiparticles strongly depends on the temperature, magnetic
field, pressure, and other parameters. The quasiparticles are well
defined and can be used to describe the scaling behavior of the
thermodynamic HF metals in a wide range of both temperatures $T$
and magnetic fields $B$.

It was demonstrated that the phase diagram of systems located near
the fermion condensation quantum phase transition are strongly
influenced by control parameters such as chemical pressure,
pressure, the number density or magnetic field. Analyzing the phase
diagram of $\rm YbRh_2Si_2$, we have found that under the
application of the chemical pressure (positive/negative) QCP is
destroyed or converted into the quantum critical line,
correspondingly.

We have analyzed the thermodynamic properties of $\rm YbRh_2Si_2$ in
both the low and the high magnetic fields. Our calculations allow us
to conclude that in magnetic field the HF system of $\rm YbRh_2Si_2$
evolves continuously without a metamagnetic transition and possible
localization of heavy $4f$ electrons. At low temperatures and under
the application of magnetic field the HF system demonstrates the LFL
behavior, at elevated temperatures the system enters the transition
region followed by the NFL behavior.

Our observations are in good agreement with experimental facts and
show that the fermion condensation quantum phase transition is
indeed responsible for the observed NFL behavior and quasiparticles
survive both high temperatures and high magnetic fields.

\section*{Acknowledgments}

We grateful to V.A. Khodel, K.G. Popov, and V.A. Stephanovich for
valuable discussions. This work was partly supported by the RFBR \#
09-02-00056.


\end{document}